\documentclass[
reprint,
superscriptaddress,
showpacs,preprintnumbers,
footinbib,
bibnotes,
amsmath,amssymb,
aps,
pra,
floatfix,
]{revtex4-2}

\usepackage{graphicx}
\usepackage{color}

\def\Jac{\textsc{Jac}}

\begin{document}

\title{Multiple Photodetachment of Silicon Anions via K-Shell Excitation and Ionization}

\author{A.~Perry-Sassmannshausen}
\affiliation{I.~Physikalisches Institut, Justus-Liebig-Universit\"{a}t Gie{\ss}en, Heinrich-Buff-Ring 16, 35392 Giessen, Germany}

\author{T.~Buhr}
\affiliation{I.~Physikalisches Institut, Justus-Liebig-Universit\"{a}t Gie{\ss}en, Heinrich-Buff-Ring 16, 35392 Giessen, Germany}

\author{M.~Martins}
\affiliation{Institut f\"{u}r Experimentalphysik, Universit\"{a}t Hamburg, Luruper Chaussee 149, 22761 Hamburg, Germany}

\author{S.~Reinwardt}
\affiliation{Institut f\"{u}r Experimentalphysik, Universit\"{a}t Hamburg, Luruper Chaussee 149, 22761 Hamburg, Germany}

\author{F.~Trinter}
\affiliation{Institut f\"{u}r Kernphysik, Goethe-Universit\"{a}t Frankfurt am Main, Max-von-Laue-Stra{\ss}e 1, 60438 Frankfurt am Main, Germany}
\affiliation{Molecular Physics, Fritz-Haber-Institut der Max-Planck-Gesellschaft, Faradayweg 4-6, 14195 Berlin, Germany}

\author{A.~M\"{u}ller}
\affiliation{Institut f\"{u}r Atom- und Molek\"{u}lphysik, Justus-Liebig-Universit\"{a}t Gie{\ss}en, Leihgesterner Weg 217, 35392 Giessen, Germany}

\author{S.~Fritzsche}
\affiliation{Helmholtz-Institut Jena, Fr{\"o}belstieg 3, 07743 Jena, Germany}
\affiliation{Theoretisch-Physikalisches Institut, Friedrich-Schiller-Universit\"{a}t Jena, 07743 Jena, Germany}

\author{S.~Schippers}
\email{stefan.schippers@physik.uni-giessen.de}
\affiliation{I.~Physikalisches Institut, Justus-Liebig-Universit\"{a}t Gie{\ss}en, Heinrich-Buff-Ring 16, 35392 Giessen, Germany}
\date{\today}

\begin{abstract}
Experimental cross sections for $m$-fold photodetachment ($m=3-6$) of silicon anions via $K$-shell excitation and ionization were measured in the photon-energy range of 1830-1900~eV using the photon-ion merged-beams technique at a synchrotron light source.  All cross sections exhibit a threshold behavior that is masked by pre-threshold resonances associated with the excitation of a $1s$ electron to higher, either partly occupied or unoccupied atomic subshells.  Results from multi-configuration Dirac-Fock (MCDF) calculations agree with the experimentally derived cross sections for photo-absorption if small energy shifts are applied to the calculated resonance positions and detachment thresholds. Moreover, a systematic approach is applied for modeling the deexcitation cascades that set in after the initial creation of a $K$-shell hole. The  resulting product charge-state distributions compare well with the measured ones for direct $K$-shell detachment but less well for resonant $K$-shell excitation. The present results are potentially useful for identifying silicon anions in cold plasmas such as interstellar gas clouds.
\end{abstract}
	
\maketitle

\section{Introduction}

Atomic anions are prototypal quantum systems for studying correlation effects since the extra electron in theses systems is bound by short-range forces.  A fundamental process that enables one  to probe correlation is photodetachment of atomic anions, i.e., the removal  of one or more electrons by absorption of a single photon. Thorough investigations of this process have repeatedly lead to a fruitful interplay between experiment and quantum theory, in particular, for outer-shell photodetachment (see, e.g., \cite{Leimbach2020a,Safronova2021} and references therein).

The correlation between core and valence electrons can be probed by inner-shell detachment where  the valence electrons are subject to strong many-electron relaxation effects following the creation of core holes \cite{Gorczyca2004a,Ivanov2004}. These effects are strongest in the case of $K$-shell detachment, which can experimentally be investigated by the photon-ion merged-beams technique using soft x-ray beams from synchrotron light sources \cite{Kjeldsen2006a,Bilodeau2012,Schippers2020a}.  So far, $K$-inner-shell photodetachment has been studied for a number of light ions comprising Li$^-$ \cite{Kjeldsen2001a,Berrah2001}, B$^-$ \cite{Berrah2007a}, C$^-$ \cite{Gibson2003a,Walter2006a,Perry-Sassmannshausen2020}, O$^-$ \cite{Gibson2012,Schippers2016a}, and F$^-$ \cite{Mueller2018b}.

Here, we extend these studies to an atomic anion where the outer shell is the $M$ shell. We present experimental cross sections $\sigma_m$ for $m$-fold photodetachment of Si$^-$ ions with $m = 3,4,5,6$ resulting in the production of multiply positively charged Si$^{(m-1)+}$ ions. This process can be represented as
\begin{equation}\label{eq:reaction}
	h\nu + \textrm{Si}^- \to \textrm{Si}^{(m-1)+} + m\, e^-.
\end{equation}
The experimental photon-energy range of 1830--1900~eV comprises the threshold for direct photodetachment of a $K$-shell electron. Furthermore, we present theoretical results for resonant and nonresonant Si$^-$ photoabsorption. Using a recently developed theoretical toolbox \cite{Fritzsche2019,Fritzsche2021},  we followed in addition the complex deexcitation cascade that sets in after the initial $K$-hole creation and gives rise to photon-energy-dependent distributions of product-ion charge states.

Apart from addressing fundamental questions in many-body physics, the present experimental and theoretical investigations are also useful for astrophysics considering the fact that negative ions have been identified, e.g., in the interstellar medium \cite{Millar2017}. In particular, they may help to improve the current astrophysical models for the x-ray absorption by silicon ions \cite{Gatuzz2020}.

\section{Experiment}\label{sec:exp}

The measurements were carried out using  the photon-ion merged-beams technique \cite{Schippers2016} at the PIPE facility \cite{Schippers2014,Schippers2020}, which is a permanently installed end-station at the photon beamline P04 \cite{Viefhaus2013} of the synchrotron radiation source PETRA\,III operated by DESY in Hamburg, Germany. Silicon anions were produced  by a Cs-sputter ion source \cite{Middleton1984} with a silicon single crystal as a sputter target and a sputter potential of about 2~kV.  {\color{black} Si$^-$ is known to posses long-lived metastable levels with excitation energies of 1.36~eV and 0.86 eV and  lifetimes of 22~s and more than 5.7~h, respectively \cite{Muell2021}. However,  Si$^-$} ions produced by a Cs-sputter ion source in an earlier experiment were found to be exclusively in their $1s^2\,2s^2\,2p^6\,3s^2\,3p^3\;^4S_{3/2}$ ground level \cite{Scheer1998}. We assume that the same holds for the present  experiment.  After acceleration to a kinetic energy of 6~keV, the ions were passed through an analyzing dipole magnet which was adjusted such that $^{28}$Si$^-$ ions were selected for further transport to the photon-ion merged-beams interaction region.
	
In the interaction region, where the residual-gas pressure was in the mid $10^{-10}$~mbar range, the ion beam was coaxially merged with the counter-propagating soft x-ray photon beam over a distance of about 1.7~m. Si$^{(m-1)+}$ ions  resulting from multiple photodetachment (Eq.~\ref{eq:reaction}) were separated from the primary ion beam by a second dipole magnet. Inside this magnet, a Faraday cup collected the primary ion beam, while the charge-selected product ions were directed to the detector chamber. Along their flight path, they first passed through a spherical 180-degree out-of-plane deflector to suppress background from stray electrons, photons, and ions and then entered a single-particle detector with nearly 100\% detection efficiency \cite{Rinn1982}.

Relative cross sections for $m$-fold photodetachment were obtained by normalizing the count rates of Si$^{(m-1)+}$ product ions on the primary Si$^-$ ion current and on the photon flux measured with a calibrated photodiode.  The ion current in the interaction region was up to 40~nA, and the photon flux amounted to $\sim 1.3\times10^{13}$~s$^{-1}$ at a photon energy spread of about 1.75~eV. The systematic uncertainty of the measured cross sections is estimated to be $\pm$15\% at 90\% confidence level \cite{Schippers2014}.

The photon-energy scale was calibrated by absorption measurements at the $2p_{1/2}$ ionization threshold of krypton in a gas-phase electron spectrometer by using the corresponding threshold energy of 1729.5~eV as measured by Kato et al.~\cite{Kato2007a} as a reference. The remaining uncertainty of the present photon-energy scale is estimated to be $\pm$1~eV. A detailed description of the general calibration procedure has been provided recently  \cite{Mueller2017,Mueller2018c}.

The experimental photon-energy bandwidth could be adjusted by choosing the exit-slit width of the photon beamline's monochromator which houses gratings with rulings of 400 lines/mm and 1200 lines/mm. We used a 1200 lines/mm grating and exit-slit widths of 1000~$\mu$m and 50~$\mu$m for low-resolution and high-resolution measurements, respectively.

\section{Computations}\label{sec:theo}

To model the photodetachment of negative ions, such as Si$^-$, detailed computations of their level structure need to be combined with an efficient treatment of the subsequent relaxation in order to account for the stepwise autoionization and photon emission of the ions. This stepwise relaxation can be formally described by an \textit{atomic cascade} that includes ions of the same element but in different charge states that are connected to each other by different (decay) processes. In this work, all calculations presented below have been performed by means of the \Jac{} toolbox, the Jena Atomic Calculator \cite{Fritzsche2019}, which supports the (relativistic) computation of atomic structures and processes. This toolbox has now been expanded recently to follow rather long ionization pathways. Here, we shall recall just a few details about such cascade computations, while the numerical results will be discussed together with the measurements in Sec.~\ref{sec:res}.

Atomic cascades often require an enormous effort to generate, combine, and simulate all data as needed for modeling a given experiment. To analyze such cascades more systematically, \Jac{} has recently been expanded for modeling different \textit{cascade schemes}, i.e., relaxation scenarios, including the production of (inner-shell) excited atoms by photoexcitation, photoionization, or other mechanisms and the stepwise decay of the excited states via photon or electron emission \cite{Fritzsche2021}. Any of these schemes starts from setting up a cascade \textit{tree}, i.e., the list of possibly contributing configurations and from selecting all relevant (decay) steps of the cascade.

In a cascade decay, each ion of a given  ensemble therefore follows an individual relaxation pathway. Obviously, a (very) large number of such pathways  occurs in the deexcitation of deep inner-shell holes owing to the different decay processes  that are possible and that have to be combined. For the stepwise decay of the $K$-shell ionized Si$^-$ ion, for example, we started from the $1s\,2s^2\,2p^6\,3s^2\,3p^3$ configuration and considered all autoionization and photon emission processes to energetically lower-lying configurations, until up to five electrons are released from the ion. For the configuration above, this results in a total of 87 configurations and about 160,000 fine-structure transitions of the Si$^{q+}\; (q=0,\ldots,5)$ atoms/ions that need to be taken into account.

To formalize the competition of the various Auger and photon emission processes, we distinguish within the \Jac{} toolbox between (so-called)  cascade \textit{computations} and \textit{simulations} \cite{Fritzsche2021}. While the cascade computations are performed with the goal to calculate and \textit{compile} the associated fine-structure Auger and photon-emission rates, the simulations then make use of these (pre-compiled) data in order to derive the ion distributions, or any other information that may be needed. This distinction has been found helpful to deal with the quite different experimental setups and observations, e.g., photon, electron, or ion distributions, that often can be  traced back to  the same (single-step) processes. \nocite{Jac-manual}

\begin{figure*}
	\includegraphics[width=\linewidth]{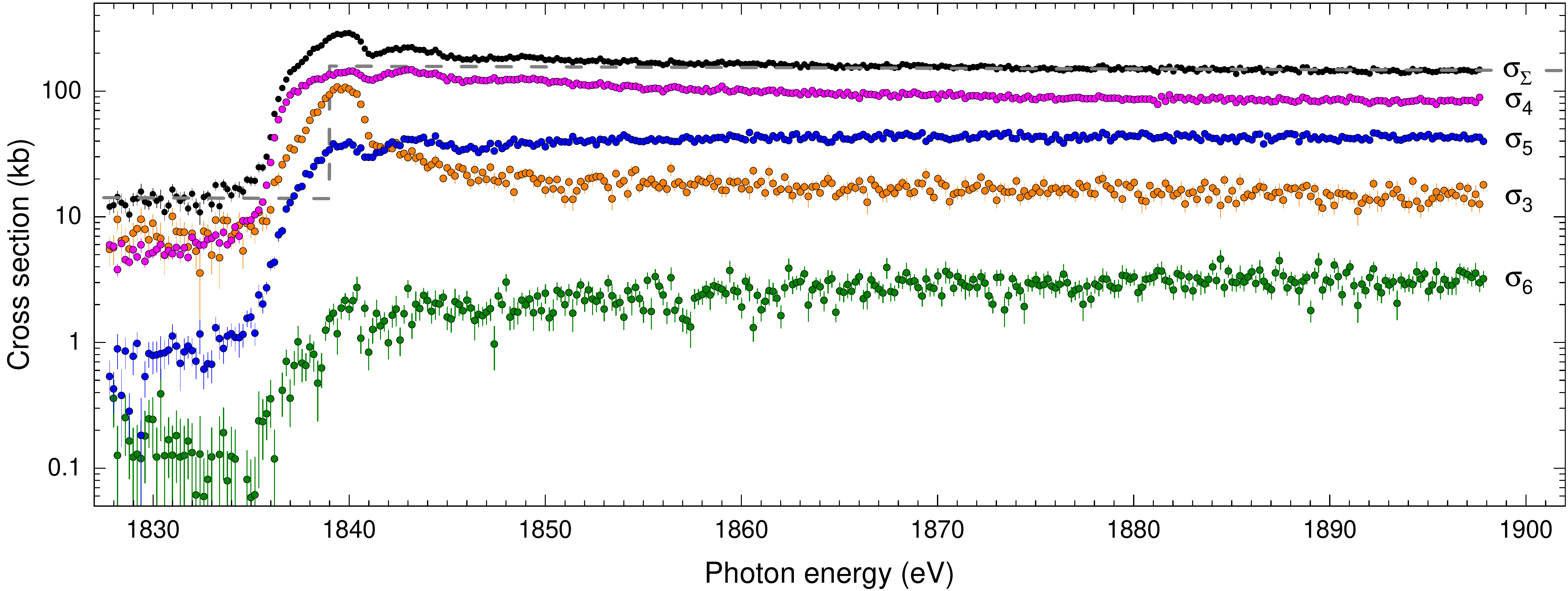}
	\caption{\label{fig:all}Overview of the measured cross sections $\sigma_m$ for multiple ($m$-fold, Eq.~\ref{eq:reaction}) detachment  of Si$^-$ ions. The differently colored data points represent $\sigma_m$  for $m$=3 (orange), $m$=4 (magenta), $m=5$ (blue), and $m=6$ (green).  The error bars only account for the statistical experimental uncertainties. The black circles represent the sum $\sigma_\Sigma$ (Eq.~\ref{eq:sum}) of all measured cross sections $\sigma_m$. The dashed grey curve is the cross section for photoabsorption of neutral silicon atoms from the compilation of Henke et al.~\cite{Henke1993}. The width of the monochromator exit slit was 1000~$\mu$m. The distorted resonance line shapes are due to a non-Gaussian contribution to the photon-energy distribution, that  becomes noticeable at large exit-slit widths for photon energies well above 1000~eV.}
\end{figure*}

Not much needs to be said about the detailed implementation of the cascade computations and simulations which has been summarized elsewhere \cite{Fritzsche2021,Jac-manual}. The major computational effort is spent on  the \textit{representation} of the fine-structure levels at each step of the cascade  as well as on  the calculation of the transition amplitudes. Within \Jac, these  are the \textit{building blocks} that enable us to formulate (and implement) all cascade computations.  Because of the complexity of atomic cascades, the (quantum-mechanical) representation of the fine-structure levels usually needs to be simplified  by \emph{tuning} the amount of the inter-electronic interactions accounted for  in the calculations. In the \Jac{} toolbox, this need for simplification is captured by an hierarchy of cascade approaches for representing all the ionic level energies and state functions. For Si$^-$, unfortunately, we were able  to apply only the simplest of these approaches, the so-called averaged single-configuration approach, in which all fine-structure states are approximated by single-configuration state functions and by just the orbitals from the initial configuration of the cascades. This approach neglects all configuration mixing between the bound-state levels and restricts the computations to just the Coulomb interaction among the electrons as well as a single set of continuum orbitals for each \textit{step} of the cascade \cite{Fritzsche1992}.

Cascade computations have been performed for the photoabsorption and photoionization cross sections in the region of the $1s \to 3p,\;4p$ inner-shell excitations and $1s$ ionization. In general, the size of such cascade computations increases rapidly  with (i) the number of electrons, that need to be replaced, and (ii) the depth of the cascade, i.e., the maximum number of electrons that can be released and, as outlined above, by the number of fine-structure transitions. This increase also results in an approximate treatment of the fine-structure representation of the ionic levels that is likely the main reason for the discrepancies of the predicted ion distributions, when compared with the measurements.

\section{Results and Discussion}\label{sec:res}

The measured low-resolution cross sections for multiple detachment of Si$^-$ are displayed in Fig.~\ref{fig:all}. All cross sections exhibit a rise at a photon energy of about 1835~eV which is associated with the resonant  excitation of a $1s$ electron to higher subshells. Two rather broad resonance features at $\sim$ 1840 and $\sim$1843~eV can be discerned in all four cross-section curves.  At higher energies, where the cross sections are dominated by nonresonant direct removal of a $K$-shell electron, they are structureless. The corresponding detachment thresholds are masked by the below-threshold resonances.

The $K$-shell hole, that is created either by resonant excitation or by nonresonant detachment, is filled by a cascade of radiative and Auger processes, which results in the observed energy-dependent distributions over the various product charge states analogous to what we have observed previously for C$^-$ \cite{Perry-Sassmannshausen2020}, where C$^{(m-1)+}$ product charge states were detected for $2\leq m \leq 6$. As compared to the C$^-$ cross sections, the present Si$^-$ cross sections are almost an order of magnitude smaller. This follows the general trend of decreasing photoionization cross sections as  the nuclear charge increases.  Another significant difference between C$^-$ and Si$^-$ concerns the dominant detachment channel which is $\sigma_2$ for C$^-$ and $\sigma_4$ for Si$^-$. {\color{black} This can be attributed to the higher number of electrons outside the $K$ shell} in the Si$^-$ ion which supports many more deexcitation channels  as compared to C$^-$. In particular, the number of possibilities for autoionization increases when going from C$^-$ to Si$^-$.

The measured relative cross sections were put on an absolute scale by matching the sum cross section
\begin{equation}\label{eq:sum}
\sigma_\Sigma = \sum_{m=3}^6\sigma_m
\end{equation}
with the  absorption cross section of neutral silicon atoms \cite{Henke1993} which is also displayed in Fig.~\ref{fig:all}. Apparently, this cross section represents the energy dependence of the experimental cross section very well except for the energies where resonances occur. In order to bring $\sigma_\Sigma$ to the same scale, all cross sections $\sigma_m$ were multiplied by the same appropriate factor. As shown explicitly for C$^-$ \cite{Schippers2020a}, where independently \emph{absolute} cross sections were experimentally determined \cite{Perry-Sassmannshausen2020}, this procedure is valid if one can safely assume that $\sigma_\Sigma$ represents all significant absorption channels. This assumption appears justified since the behavior of the cross sections $\sigma_m$
suggests that the unmeasured single and double detachment channels and the higher detachment channels with $m>6$ are negligible. The single-detachment channel could not be measured since our experimental setup is only capable of measuring charged product ions. Moreover, the measurement of the (weak) double-detachment channel was severely hampered by a large background from collisions of Si$^-$ with the residual gas such that meaningful results could not be obtained within a reasonable amount of time. Measurements of the multiple-detachment cross sections for $m>6$ were not attempted.

\begin{figure}
 \includegraphics[width=\linewidth]{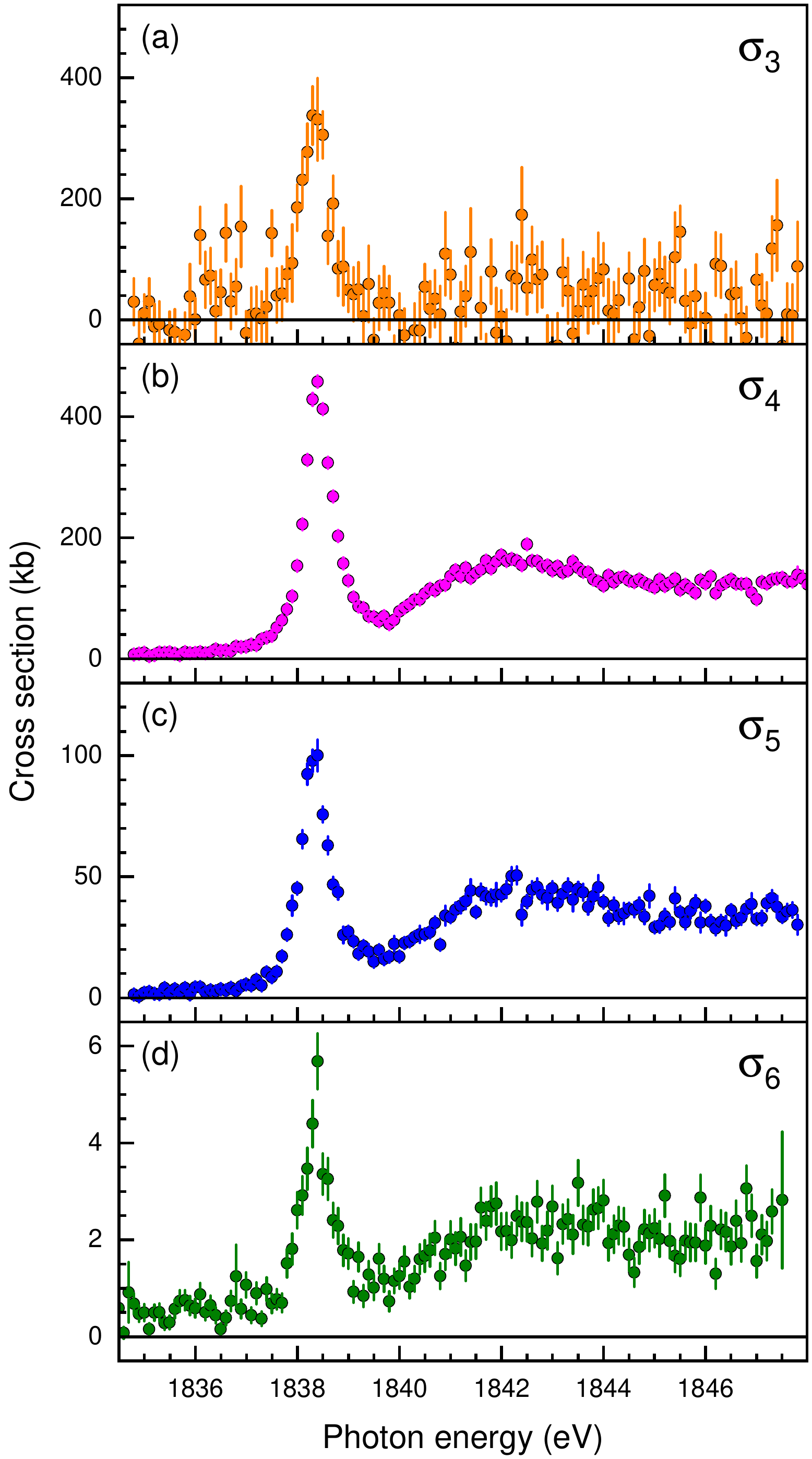}
	\caption{\label{fig:all50mu}High-resolution cross sections $\sigma_m$ for $m$-fold ($m$=3-6) photodetachment of Si$^-$ ions in the energy region of the near-threshold resonances. The width of the monochromator exit slit was 50~$\mu$m.}
\end{figure}	

The nonzero cross sections below  1835~eV are due to direct detachment of an $L$-shell electron and the subsequent (multiple) autoionization. In this energy range, the sum cross section $\sigma_\Sigma$ agrees, as mentioned above, well with the cross section for photoabsorption of neutral silicon atoms. This is also the case for energies above 1855~eV which are beyond the threshold for the direct detachment of a $K$-shell electron. For neutral silicon this threshold has been predicted to be at 1838.9~eV \cite{Henke1993} (dashed line in Fig.~\ref{fig:all}). For Si$^-$, the corresponding step-like cross-section features are masked by the resonances that occur immediately below these thresholds.

In order to obtain more detailed experimental  information on the photodetachment  resonances we performed additional cross-section measurements with higher  resolution as compared to the measurements shown in Fig.~\ref{fig:all}. These high-resolution measurements are displayed in Fig.~\ref{fig:all50mu}. The statistical uncertainties of $\sigma_3$ are comparatively large because a substantial background contributed to the measured signal in this channel. Such a background was largely absent in the other scrutinized channels. In all measured cross sections, the strongest feature is a comparatively narrow resonance at $\sim$1838.4~eV with an experimental width of  0.55~eV. A second, broader resonance feature at about 1842.4~eV with a width of 2--3~eV is visible in the cross sections for fourfold, fivefold, and sixfold detachment. The relative strength of this feature, when compared to the narrower one,  increases with increasing product-ion charge state.

\begin{figure}[t]
	\includegraphics[width=\columnwidth]{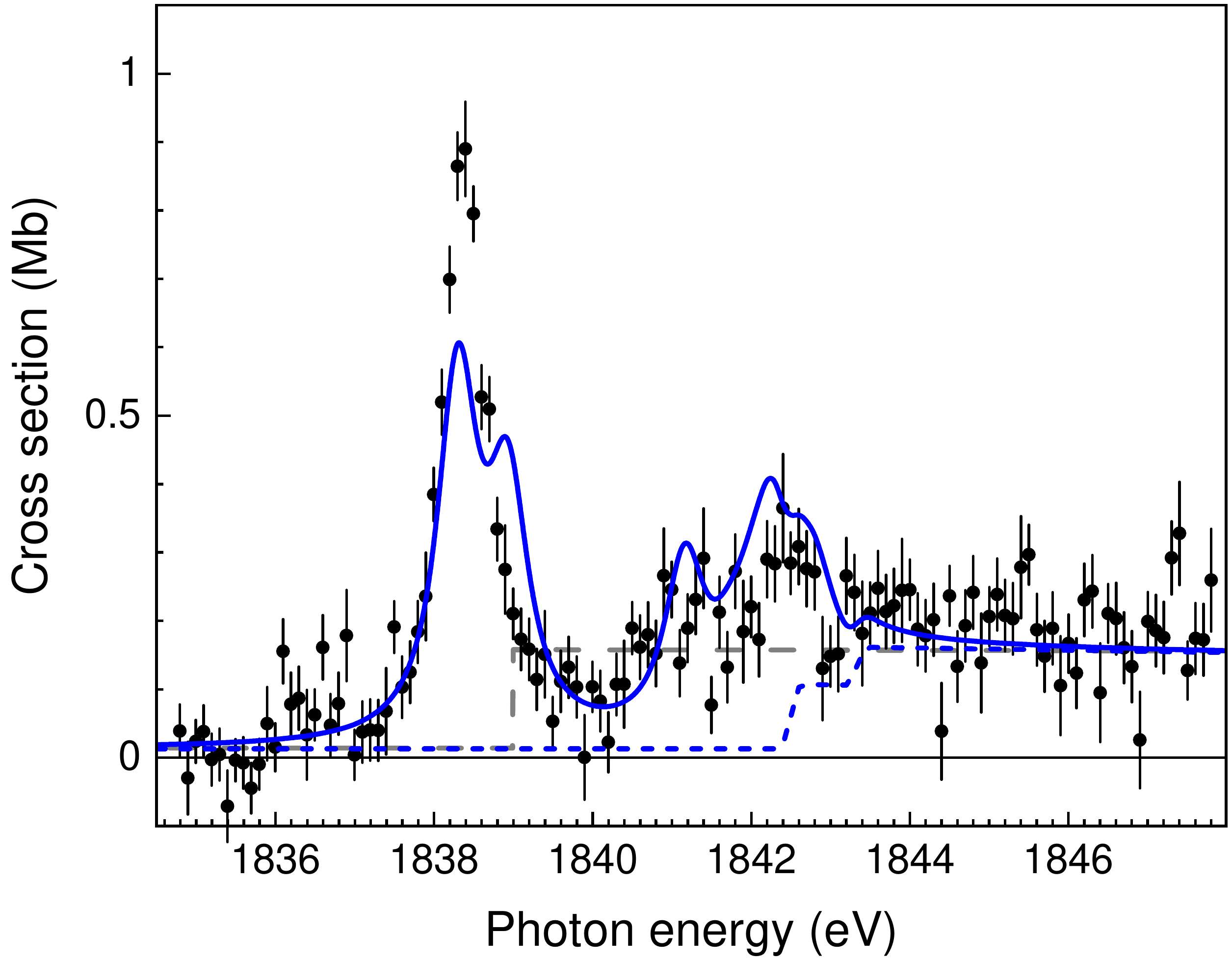}
	\caption{\label{fig:theo50}Comparison between the high-resolution experimental sum cross section $\sigma_\Sigma$ (sum of the  cross sections displayed in Fig.~\ref{fig:all50mu}) with the present theoretical absorption cross section (full line). The short-dashed line represents the contribution by direct detachment of a single $L$-shell or $K$-shell electron to the absorption cross section. As in Fig.~\ref{fig:all}, the long-dashed  curve is the cross section for photoabsorption of neutral silicon atoms \cite{Henke1993}.}
\end{figure}

A double-peak structure was similarly observed  in the multiple photodetachment of C$^-$ \cite{Perry-Sassmannshausen2020}, where the first resonance was attributed to a $1s\to2p$ excitation to the $1s\,2s^2\,2p^4\;^4P$ term and the second, broader resonance peak was attributed to higher excitations to $^4P$ terms mainly of the $1s\,2s^2\,2p^3\,3p$ configuration. In analogy, one might therefore assign the $1s^{-1}\,3p^4\;^4P$ term to the first $K$-shell excited resonance of Si$^-$ and a blend of several $^4P$ terms associated with $1s\to4p$ excitations  to the second resonance peak. Our calculations confirm this expectation,  though one should keep in mind that they only provide a coarse orientation on the positions. For any negative ion, the calculated levels result from strong configuration interactions.

Figure \ref{fig:theo50} shows the comparison between the sum of the experimental cross sections from Fig.~\ref{fig:all50mu} and our theoretical absorption cross section.  The individual theoretical resonances are represented by  Voigt line profiles with a Gaussian width of 0.05 eV that accounts for the experimental photon-energy spread.  The calculated Lorentzian width of virtually all these resonances amounts to 0.57~eV. It is determined mainly  by the KLL Auger transition that fills the core hole and it is about 15\% larger than the  $K$-level width of neutral silicon \cite{Campbell2001}. An energy shift of $-1.3$~eV was applied to the theoretical resonance positions in order to line these up with the experimental peak positions.  This shift is within the uncertainty of the theoretical resonance energies. In order to achieve the best possible agreement  between theory and experiment, a shift of similar magnitude was applied to the calculated threshold energies for direct $K$-shell detachment resulting in the values 1842.5 and 1843.3~eV for the $1s^{-1}\,3p^3\,(^4S)\;^5S_2$ and $1s^{-1}\,3p^3\,(^4S)\;^3S_1$ levels in neutral silicon, respectively.  Apart from some obviously not quite correct relative line strengths, the general agreement between experiment and theory is very satisfying considering the complexity of the problem under study. Furthermore, we like to point out that, at energies above $\sim$1846 ~eV,   our theoretical cross section for photoabsorption of Si$^-$ agrees excellently with the absorption cross section for photoabsorption of neutral silicon \cite{Henke1993}. This supports our earlier assumption that the presence of the extra electron in Si$^-$ does not influence the cross section for direct inner-shell detachment.

\begin{figure}[t]
	\includegraphics[width=\columnwidth]{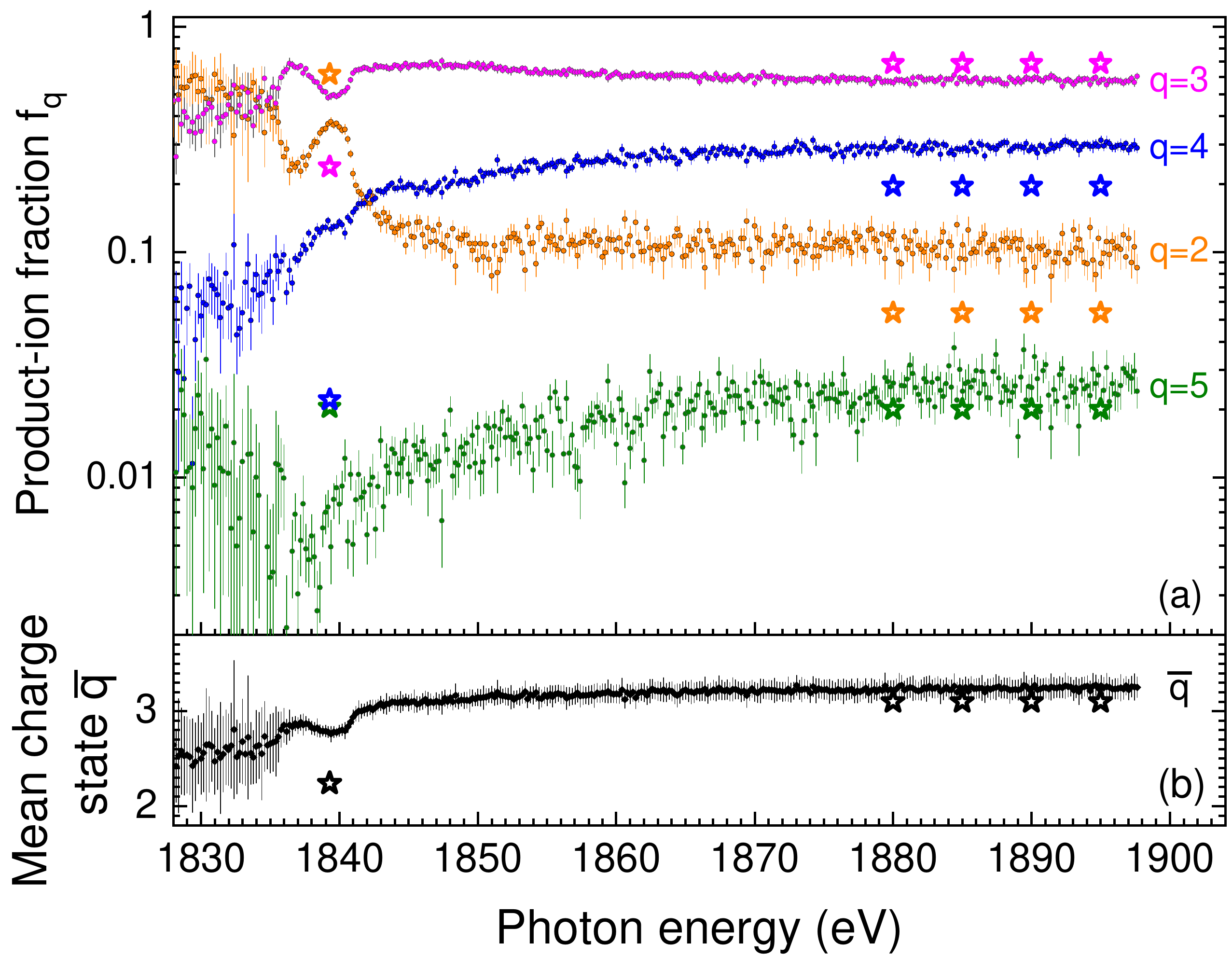}
	\caption{\label{fig:frac}Experimental (small filled circles with error bars) and theoretical (large stars) Si$^{q+}$ product-ion fractions (Eq.~\ref{eq:fq}, panel a) and mean charge state $\overline{q}=\sum_qqf_q$ (panel b).}
\end{figure}

In Fig.~\ref{fig:frac}(a), the outcome of our cascade calculations, i.e., the distribution of product-ion charge states $q=m-1$ resulting from $m$-fold photodetachment (stars in Fig.~\ref{fig:frac}), is compared to the corresponding experimental values
\begin{equation}\label{eq:fq}
f_q = \sigma_{q+1}/\sigma_\Sigma.
\end{equation}

{\color{black}The behavior of these charge-state fractions can be qualitatively understood as follows. The cross sections below $\sim$1837~eV, i.e., below the first $K$-shell excitation energy are mainly due to $L$-shell detachment and the subsequent autoionization that fills the $L$-shell hole. Over the narrow displayed energy range below 1837~eV, the branching ratios for the production of the various final charge states are virtually constant. Therefore, all partial cross sections $\sigma_{q+1}$ have a similar progression as functions of photon energy and, thus, also  the fractions $f_q$. This changes when a $K$-shell hole is created by excitation or direct detachment. The subsequent Auger processes that fill the $K$-shell hole significantly influence the branching ratios giving higher weights to the higher charge states. This is apparent for the $1s\to3p$ resonance at about 1838~eV where the fraction $f_3$ rises to above the fraction $f_2$.  A similar behavior also occurs (less drastically) at  the $K$-shell detachment-threshold energy of $\sim$1841~eV.}

At energies above $\sim$1880~eV, the distribution results {\color{black} almost exclusively} from direct detachment of a $K$-shell electron and the subsequent Auger cascade that produces the various product-ion charge states. The calculated fractions are of the correct orders of magnitude. For individual charge states the theoretical fractions deviate from the corresponding experimental values by factors of up to 2. However, these uncertainties average out when the mean product-ion charge state is considered [Fig.~\ref{fig:frac}(b)].

 {\color{black}We attribute the discrepancies between the experimental and theoretical $f_g$ values to the approximations that we had to apply for making the cascade computations tractable (see also Sec.~\ref{sec:theo}). The product-ion distributions in our computations are quite strongly affected by the thresholds for the subsequent Auger processes. Because  we could only accommodate an averaged single-configuration approach, the fine structure is poorly taken into account and does not describe all the ionization channels properly. Very likely, this is the main reason that the theoretical predictions for $f_2$ and $f_4$ are underrated. In the future, we hope to advance our cascade simulations, although this will require some major developments and tests for some simpler system with, say, $Z < 10$. In the present cascade model, the ion fractions $f_0$ and $f_1$ are very tiny and cannot be compared directly with the experimental data. Moreover, the current cascade model does not predict any ions in the $q\geq 6$ channels. More accurate predictions for the higher product-ion charge states will also require the incorporation of simultaneous double-detachment processes, either during the initial detachment and/or the subsequent decay (see, e.g., \cite{Beerwerth2019,Mueller2021b}).}

Below the threshold for $K$-shell detachment, the distribution of product charge states depends on the individual $K$-shell excited core-hole levels. Computationally, this is even more demanding than the calculation of the cascades that follow the direct $K$-shell detachment, since the core-excited configurations then contain one more electron as compared to the core-ionized configurations. We have performed one such calculation for the $1s^{-1}\,3p^4$ configuration. The resulting product-ion fractions (stars at $\sim$1839 eV in Fig.~\ref{fig:frac}) compare less well with the experimental ones. The fractions $f_3$ and $f_4$ are underestimated by more than a factor of 2 resulting in a significantly too small mean charge state.

\section{Summary and Conclusions}

The present joint experimental and theoretical study of multiple photodetachment of the silicon anion via $K$-shell excitation and ionization extends  our previous work with light ions (see \cite{Schippers2020a} and references therein) to an atomic system with a partly occupied $M$ shell. This additional valence shell challenges experiment and theory alike.  The experimental challenge arises from the fact that $K$-shell detachment cross sections decrease with increasing nuclear charge. The cross section for the direct detachment of a $1s$ electron is an order of  magnitude smaller for Si$^-$ as compared to C$^-$. The photon-ion merged-beams method as implemented at the PETRA\,III synchrotron successfully proved capable of meeting this challenge, with the high photon flux from the XUV beamline P04 and  the high product selectivity of the PIPE setup being the key assets \cite{Schippers2020} .

The present theoretical calculations were mainly challenged by the strong correlation effects that govern all anions and by the complexity of the deexcitation cascades following after the initial creation of a $K$-shell vacancy. In particular, the presence of the $M$-shell places additional demands on the cascade calculations as compared to our earlier studies with light negative ions that have been summarized in Ref.~\cite{Schippers2020a}.  The large amount of levels and transitions to be followed in such calculations requires a systematic and consistent approach as now offered by the recently developed \Jac{} toolbox \cite{Fritzsche2019,Fritzsche2021}.

The measured cross sections for net triple to sixfold photodetachment exhibit a threshold for direct photodetachment of one $K$-shell electron. This threshold  is masked by resonances associated with $1s\to3p$ and $1s\to4p$ excitations in the language of single-particle excitations. The calculated resonance positions agree with the values from a high-resolution measurement by less than 2 eV.  While the calculated relative resonance strengths agree reasonably well with our measurements, they leave room for improved  theoretical approaches. The same is true for the fractions of the product ions as obtained from the cascade computations. These turn out to be more reliable for the deexcitation cascade that follows direct $K$-shell detachment than for the one that evolves after $1s\to3p$ excitation.

The statistical quality of the present data is somewhat lower as compared to a recently published and similarly comprehensive study on multiple photodetachment of C$^-$ ions \cite{Perry-Sassmannshausen2020}. This is partly due  to the already mentioned reduced cross section but also due to a lower photon flux which varies with photon energy such that it rapidly decreases at energies above 1000~eV.  Since C$^-$([He] $2s^2\,2p^3$) and Si$^-$([Ne] $3s^2\,3p^3$) have analogue open-shell structures, one might expect similarities in the progression of the detachment cross sections. Indeed, the double-resonance structure at the $K$-shell detachment threshold is common to both species.  However, additional  small resonances at higher energies, that were discovered in the C$^-$ triple-detachment channel, could not be detected in any of the measured Si$^-$ photodetachment channels within the present  level of statistical uncertainty. Finally, we would like to mention that the experimentally derived and calculated cross sections for Si$^-$ photoabsorption might be useful for identifying silicon anions in cold cosmic gas clouds by their characteristic x-ray absorption pattern.
	
\begin{acknowledgments}
We acknowledge DESY (Hamburg, Germany), a member of the Helmholtz Association HGF, for the provision of experimental facilities. Parts of this research were carried out at PETRA\,III and we would like to thank  Kai Bagschik, Frank Scholz, Jörn Seltmann, and Moritz Hoesch for assistance in using beamline P04. We are grateful for support from Bundesministerium f\"{u}r Bildung und Forschung within the \lq\lq{}Verbundforschung\rq\rq\ funding scheme (grant nos.\ 05K19GU3 and 05K19RG3) and from Deutsche Forschungsgemeinschaft (DFG, project no.\  389115454). 
\end{acknowledgments}


%
\end{document}